\documentstyle[12pt]{article}
\catcode`\@=11
\def\marginnote#1{}
\def\ifmath#1{\relax\ifmmode #1\else $#1$\fi}

\def\bold#1{\setbox0=\hbox{$#1$}%
     \kern-.025em\copy0\kern-\wd0
     \kern.05em\copy0\kern-\wd0
     \kern-.025em\raise.0433em\box0 }

\def\GENITEM#1;#2{\par\vskip6pt \hangafter=0 \hangindent=#1
   \Textindent{$ #2$ }\ignorespaces}

\newcount\hour
\newcount\minute
\newtoks\amorpm
\hour=\time\divide\hour by60
\minute=\time{\multiply\hour by60 \global\advance\minute by-
\hour}
\edef\standardtime{{\ifnum\hour<12 \global\amorpm={am}%
    \else\global\amorpm={pm}\advance\hour by-12 \fi
    \ifnum\hour=0 \hour=12 \fi
    \number\hour:\ifnum\minute<100\fi\number\minute\the\amorpm}}
\edef\militarytime{\number\hour:\ifnum\minute<100\fi\number\minute}
\def\draftlabel#1{{\@bsphack\if@filesw {\let\thepage\relax
  \xdef\@gtempa{\write\@auxout{\string
    \newlabel{#1}{{\@currentlabel}{\thepage}}}}}\@gtempa
    \if@nobreak \ifvmode\nobreak\fi\fi\fi\@esphack}
     \gdef\@eqnlabel{#1}}
\def\@eqnlabel{}
\def\@vacuum{}
\def\draftmarginnote#1{\marginpar{\raggedright\scriptsize\tt#1}}
\def\draft{\oddsidemargin -.5truein
        \def\@oddfoot{\sl preliminary draft \hfil
        \rm\thepage\hfil\sl\today\quad\militarytime}
        \let\@evenfoot\@oddfoot \overfullrule 3pt
        \let\label=\draftlabel
        \let\marginnote=\draftmarginnote

\def\@eqnnum{(\theequation)\rlap{\kern\marginparsep\tt\@eqnlabel}%
\global\let\@eqnlabel\@vacuum}  }
\def\preprint{\twocolumn\sloppy\flushbottom\parindent 1em
        \leftmargini 2em\leftmarginv .5em\leftmarginvi .5em
        \oddsidemargin -.5in    \evensidemargin -.5in
        \columnsep 15mm \footheight 0pt
        \textwidth 250mmin      \topmargin  -.4in
        \headheight 12pt \topskip .4in
        \textheight 175mm
        \footskip 0pt

\def\@oddhead{\thepage\hfil\addtocounter{page}{1}\thepage}
        \let\@evenhead\@oddhead \def\@oddfoot{} \def\@evenfoot{}
}
\def\titlepage{\@restonecolfalse\if@twocolumn\@restonecoltrue\onecolumn
     \else \newpage \fi \thispagestyle{empty}\c@page\z@
        \def\thefootnote{\fnsymbol{footnote}} }
\def\endtitlepage{\if@restonecol\twocolumn \else  \fi
        \def\thefootnote{\arabic{footnote}}
        \setcounter{footnote}{0}}  
\catcode`@=12
\relax

\def\be{\begin{equation}}
\def\ee{\end{equation}}
\def\bea{\begin{eqnarray}}
\def\eea{\end{eqnarray}}
\def\simlt{\stackrel{<}{{}_\sim}}
\def\simgt{\stackrel{>}{{}_\sim}}

\def\NPB#1#2#3{{\it Nucl.~Phys.} {\bf{B#1}} (19#2) #3}
\def\PLB#1#2#3{{\it Phys.~Lett.} {\bf{B#1}} (19#2) #3}
\def\PRD#1#2#3{{\it Phys.~Rev.} {\bf{D#1}} (19#2) #3}
\def\PRL#1#2#3{{\it Phys.~Rev.~Lett.} {\bf{#1}} (19#2) #3}
\def\ZPC#1#2#3{{\it Z.~Phys.} {\bf C#1} (19#2) #3}

\def\mst11{m_{\;\widetilde{t}_{1}}}

\def\mst22{m_{\;\widetilde{t}_{2}}}
\def\mst12{m_{\;\widetilde{t}_{1,2}}}

\def\msb11{m_{\;\widetilde{b}_{1}}}
\def\msb22{m_{\;\widetilde{b}_{2}}}
\def\msb12{m_{\;\widetilde{b}_{1,2}}}

\def\mtilde2{\widetilde{m}^{2}}

\relax

\begin{document}
\topmargin-2.5cm
%
\begin{titlepage}
\begin{flushright}
FNAL-PUB/97-XXX\\
CERN-TH/97-373\\
hep--ph/9801251 \\
\end{flushright}
\vskip 0.3in 
\begin{center}{\Large\bf
Photon signatures for low energy supersymmetry breaking
and broken R-parity} 
\vskip .7 cm
{\bf M. Carena~$^{a}$},
{\bf S. Pokorski~$^{b,c}$} and {\bf C.E.M. Wagner~$^{b}$}
\vskip.35in
$^a$~Fermi National Accelerator Laboratory, P.O. Box 500, Batavia, 
IL 60510, U.S.A.
$^{b}$~CERN, TH Division, CH--1211 Geneva 23, Switzerland\\
$^{c}$~Institute of Theoretical Physics, Warsaw University, Poland. 
\end{center}
\vskip1.3cm
\begin{center}
{\bf Abstract}
\end{center}
\begin{quote}
The possible phenomenological consequences of R-parity violating interactions
in the framework of low energy supersymmetry breaking are studied.  
It is pointed out that even very weak R-parity violation would completely
overshadow one of the basic signatures of low energy supersymmetry 
breaking models, that is the decay of the next to 
lightest supersymmetric particle into a photon (lepton) and missing
energy. Thus, the observation of these decays would put very strong
limits on R-parity violating couplings. Vice-versa, if R-parity violation
is established experimentally, before a detailed knowledge of the spectrum
is obtained,
it will be very difficult to distinguish gravity mediated from 
low energy gauge mediated supersymmetry breaking scenarios. Those
conclusions are very model independent. We also comment on the possibility
of mixing between charged and neutral leptons with charginos and
neutralinos, respectively, and its phenomenological consequences
for the photon (lepton) signatures, in scenarios where this mixing
is generated by the presence of bilinear or trilinear  R-parity
violating terms in the superpotential.
\end{quote}
\vskip1.cm
\begin{flushleft}
CERN-TH/97-373\\
December 1997 \\
\end{flushleft}
\end{titlepage}
\setcounter{footnote}{0}
\setcounter{page}{0}
\newpage
%

There are two basic open questions in supersymmetric models: The
mechanism of supersymmetry (SUSY) breaking and the theory of flavor.
The latter includes in particular the question about
R-parity conservation or the pattern of its breaking. The two questions
may or may not be correlated. In models in which supersymmetry
breaking is transmited to the observable sector via gauge 
interactions at energies much lower than the grand unification scale,
the flavor physics is most likely  
decoupled from the mechanism of supersymmetry breaking, due to a 
large separation in the relevant mass scales. The gauge mediation
of SUSY breaking  and R-parity violation is certainly one of the
possible scenarios. The signatures for
both, low energy gauge mediated supersymmetry breaking and R-parity
violation are based on rare decays and it is hence of great importance
to determine if both  ideas can be independently established 
experimentally. 

A quite general signature for 
low energy gauge mediated scenario of supersymmetry
breaking is the decay of the next to
lightest supersymmetric particle (NLSP) into
gravitinos (the actual lightest supersymmetric particle, in this
case)~\cite{Fayet},
\begin{equation}
\tilde{\chi}^0_1 \rightarrow \gamma \; \tilde{G}; \;\;\;\;\;\;\;
{\rm or}  \;\;\;\;\;\; \tilde{l} \rightarrow l \; \tilde{G},
\label{eq:1}
\end{equation}
depending on which sparticle is the NLSP.\footnote{If the $\tilde{\chi}^0_1$ 
is mainly Higgsino- like, then the decay
$\tilde{\chi}^0_1 \rightarrow h \; \tilde{G}
\rightarrow b \bar{b} \tilde{G}$,  where $h$ is the lightest Higgs 
boson (but can also be the heavy CP-even Higgs 
or the CP-odd Higgs boson if
they are sufficiently light), is also possible.}
The decay rate, Eq(\ref{eq:1}), is given
by \cite{gms}-\cite{at2},
\begin{equation}
\Gamma  \simeq K \frac{m_{NLSP}}{48 \pi} \left(\frac{m_{NLSP}}
{\sqrt{M_P m_{\tilde{G}}}}\right)^4, 
\label{rategm}
\end{equation}
where $M_P$ is the Planck mass, $m_{NLSP}$ is the mass of the NLSP
and $K$ is  a projection factor equal to the square of the component 
in the NLSP of the
superpartner of the particle the NLSP is decaying into. For instance,
if the NLSP is a neutralino, which decays into a photon  and a gravitino,
$K$ is just the square of the 
photino component of the lightest neutralino, which
is given by
\begin{equation}
K = \left| N_{11} \cos\theta_W + N_{12} \sin\theta_W \right|^2,
\end{equation}
where $N_{ij}$ is the mixing matrix connecting the 
neutralino mass eigenstates
to the weak eigenstates in the basis $\tilde{B}, \tilde{W}, \tilde{H}_1,
\tilde{H}_2$.
The decay rate is inversely proportional to the square of the gravitino 
mass, which is given by
\begin{equation}
m_{\tilde{G}} M_P \simeq F_{SUSY}/\sqrt{3},
\end{equation}
where $\sqrt{F_{\rm SUSY}}$ is the supersymmetry breaking scale.

For our purpose, most interesting are the limits on $F_{SUSY}$ coming
from the requirement of detectability, this means, that the NLSP decays 
before escaping the detector. From Eq. (\ref{rategm}),
an upper bound on
the supersymmetry breaking scale,
$\sqrt{F_{\rm SUSY}} \simlt 10^7$ GeV, may be obtained from the requirement
that the NLSP decays into gravitinos within the detector \cite{at2}.
Low energy  SUSY breaking scenarios,
with observable decays into photon and gravitino will be characterized
by events containing photons and missing energy, in contrast to 
supergravity scenarios, where, unless very specific conditions are
fulfilled~\cite{radneu}, \cite{m2eqm1}, 
photons do not represent a characteristic signature.

Renormalizable R-parity violating scenarios 
are well known \cite{hall}. They
are derived from the effective low-energy superpotential
\begin{equation}
W = \lambda_{ijk} L_i L_j \bar{E}_k + \lambda^{'}_{ijk} L_i Q_j \bar{D}_k
+ \lambda''_{ijk} \bar{U}_i \bar{D}_j \bar{D}_k + \epsilon_i L_i H_2,
\label{eq:superpot}
\end{equation}
where $L$ ($E$) are the left- (right-) handed lepton $SU(2)_L$
doublet (singlet) superfields, $Q$ are the left handed quark 
doublet superfields,
and $U$ and $D$ are the up and down right-handed quark superfields.
If present, R-parity violating couplings may induce a large
number of new lepton or baryon number violating processes. Hence,
strong experimental limits on many of the R-parity violating
couplings, or on their products, can be set \cite{rpreview}. 
We discuss trilinear
and bilinear terms independently and begin with the former.

One obvious, but important, observation is that,
in the case of a neutralino as the NLSP, 
the presence of any
single trilinear R-parity violating coupling would induce the decay
\begin{equation}
\tilde{\chi}^0_1 \rightarrow f f f .
\end{equation}

In the case of lepton violating couplings and sleptons as the
NLSP,
\begin{equation}
\tilde{l} \rightarrow ff,
\end{equation}
where $f$ are fermionic states, which may be leptonic or hadronic
depending on the particular cases. 
The rate for the decay of neutralino into fermions 
via R-parity violating interactions reads \cite{hall},\cite{Dawson}
\begin{equation}
\Gamma(\tilde{\chi}^0_1 \rightarrow f \; f \; f )
\simeq \frac{\lambda^2 \; \alpha_w}{128 \pi^3} 
\left[ N_{11} \tan\theta_w \frac{Y_f}{2} + N_{12} T^3_f \right]^2
N_C^{f}  
\frac{m_{\tilde{\chi}^0_1}^5}{m_{\tilde{f}}^4},
\label{threeb}
\end{equation}
where $\lambda$ is the general R-parity violating coupling, 
$m_{\tilde{f}}$ is the mass of the light sfermion 
interacting via the R-parity violating coupling, $N_C^f$, $Y_f$
are  
its color number and hypercharge, respectively, and
a relevant gaugino component of the NLSP has been assumed.
The R-parity violating two-body  decay of a slepton into 
fermions reads
\begin{equation}
\Gamma(\tilde{l} \rightarrow f f) 
\simeq \lambda^2 \frac{m_{\tilde{l}}}{16 \pi}.
\end{equation} 

It is interesting to compare 
these decay
rates with the rates of decay for low energy supersymmetry
breaking models,
Eq. (\ref{rategm}), as a function of the coupling 
$\lambda$ and of the NLSP mass and the mass of the virtually exchanged
sparticle. 
The comparison  is striking. In the case of
neutralinos as the NLSP,
if photon events
are to be seen in the presence of R-parity violating couplings,
it follows from Eq. (\ref{threeb}) that
the coupling must be bounded by
\begin{equation}
\lambda \simlt 100 \left( \frac{m_{\tilde{f}}}{\sqrt{F_{\rm SUSY}}} 
\right)^2 , 
\label{bound1}
\end{equation}
which is a very stringent bound. Actually, for  values of
the sfermion masses, $m_{\tilde{f}} \simeq 300$ GeV, 
and values of $\sqrt{F_{\rm SUSY}} \simeq 10^6$ GeV, which are
typical of low energy gauge mediated supersymmetry breaking 
models, the bound on the coupling is given by
\begin{equation} 
\lambda \simlt   10^{-5}. 
\end{equation}
This bound is of the order of magnitude of the strongest bounds
on the trilinear R-parity violating couplings coming from flavor, lepton
and baryon number violating processes \cite{rpreview}.
Such low values of the couplings imply that R-parity violation
is unlikely to be seen experimentally.

There is one exception to this general conclusion. This is the
case in which the mainly gaugino-like neutralino is the NLSP and 
the dominant R-parity violating coupling
is baryon violating, involving the third generation
right handed top superfield, i.e. $\lambda''_{3ij}$.
In this case, assuming that the neutralino is lighter than the
top quark, due to phase space constraints 
the decay of the lightest neutralino into three fermions
will proceed via the small mixing between the stop and scharm eigenstates
(which can be present at the tree level or appear via the RG evolution):
\begin{equation}
\tilde{\chi}^0_1 \rightarrow c j j.
\end{equation}
In case the mixing is induced by radiative corrections, 
the additional loop-factors
\cite{hikasa} 
will suppress the R-parity
violating decay. Due to those
extra supression factors the bounds on $\lambda''$
will be  more than two orders of magnitude weaker than in
Eq. (\ref{bound1}). Even in this case, the bound is 
stringent, and hence, the observation of photons in the final
state will still require  small values of this R-parity violating
coupling. Observe that an analogous suppression factor would  not appear
if $\lambda'_{3ij}$ were the dominant R-parity violating coupling, 
since in this case the neutralino can 
decay into left-handed bottoms without any phase space suppression.

In the case of sleptons as NLSP, the bound is even more striking,
\begin{equation}
\lambda \simlt \left( \frac{m_{\tilde{f}}}{\sqrt{F_{\rm SUSY}}} \right)^2,
\label{bound2}
\end{equation}
meaning that the bound is typically two orders of magnitude 
smaller than in the case of neutralino as the lightest supersymmetric
particle. 

In the
neutralino case,
since all sfermions tend to couple to the
NLSP, the bounds are quite general, on any possible $\lambda$.
The strong bound on the couplings in the sfermion NLSP case,
instead, only applies to the couplings of the lightest sfermion.
Larger coupling of the heavier sfermions do not interfere with the
possible signatures characteristic of the gauge mediated
scenario. 
In general, assuming that the bound on the R-parity violating
coupling of the NLSP, Eq. (\ref{bound2}), is fulfilled, 
two possibilities may occur:
{\it i)} The other
R-parity violating couplings are sufficiently
strong, so that two body R-parity violating decays of non-LSP
sfermions can be observed; {\it ii)} Either the R-parity violating 
couplings for
most of the sfermions are small and/or the neutralino is light 
enough so that the dominant decay mode for non-NLSP sfermions is
R-parity conserving.

In 
case {\it i)} it would be easy to have independent experimental tests of R-parity
violation and low energy supersymmetry breaking. 
In order to get a quantitative estimate of the values of the
R-parity violating couplings for which this occurs, let us
consider the R-parity conserving decay of a sfermion into
a fermion and a neutralino:
\begin{equation}
\tilde{f} \rightarrow f \tilde{\chi}^0_1; \;\;\;\;\;\;\;\;\;
\tilde{\chi}^0_1 \rightarrow l + NLSP \rightarrow l \bar{l} \tilde{G}.
\end{equation}
In the above, the lightest neutralino,
$\tilde{\chi}^0_1$ may be real or virtual.
If the neutralino is lighter than the sfermion, 
the sfermions may decay into
neutralinos via a two-body decay channel, with a rate
\begin{equation}
\Gamma( \tilde{f} \rightarrow f \tilde{\chi}^0_1 ) \simeq
\frac{g^2}{8 \pi} \left( N_{12} T_f^3 + \frac{Y_f}{2} \tan\theta_W
N_{11} \right)^2
m_{\tilde{f}}
\left( 1 - \frac{m_{\tilde{\chi}^0_1}}{m_{\tilde{f}}} \right)^2,
\end{equation}
where 
we have ignored terms proportional to the fermion Yukawa coupling.
Due to the largeness of the weak gauge couplings, the R-parity
violating decays are dominant only if
\begin{equation}
\lambda> 0.1,
\label{bound11}
\end{equation}
unless
cancellations between the different contributions 
to the effective sfermion-fermion-neutralino coupling take place, or
the neutralino mass is close to the sfermion mass. 
Such large R-parity violating couplings are subject to strong
experimental constraints, and can only be realized in some
particular cases \cite{HERA}.

Even if the neutralino were heavier than the sfermion, the dominance of
the R-parity violating decay modes may still require fairly
large couplings $\lambda$, due to
the existence of three body R-parity conserving decay channels.
For instance, let us consider the three body decay of a sfermion
$\tilde{f}_i$ 
into another one, $\tilde{f}_j$, considered to be the NLSP.
Assuming 
$m_{\tilde{\chi}^0_1}^2,
m_{\tilde{f}_i}^2 \gg m_{\tilde{f}_j}^2$, we get~\cite{CGLW}
\begin{equation}
\Gamma (\tilde{f}_i \to f_i \bar{f}_j \tilde{f}_j) \simeq
\frac{g^4}{48(2\pi)^3}
m^3_{\tilde{f}_i}  
\sum_{a,b=1}^4 
\frac{A_a^i A_b^j}{m_{\tilde{\chi}^0_a}m_{\tilde{\chi}^0_b}} ~,
\end{equation}
\newline
and
\newline
\begin{equation}
\Gamma (\tilde{f}_i \to f_i f_j \bar{\tilde{f}_j}) \simeq
\frac{g^4}{192(2\pi)^3}
m^5_{\tilde{f}_i}  
\sum_{a,b=1}^4 
\frac{A_a^i A_b^j}{m_{\tilde{\chi}^0_a}^2 \; m_{\tilde{\chi}^0_b}^2} ~,
\end{equation}
where
\begin{equation}
A_a^i\equiv \left( N_{a1} \frac{Y_{f_i}}{2} \tan\theta_W +
N_{a2} T^3_{f_i}\right)
\end{equation}
and we are ignoring factors of order one and the phase space factors.
We have implicitly assumed that $f_i$ and $f_j$ have opposite
chiralities.
Observe the different behaviour with the neutralino
masses of the fermion conserving and fermion violating processes,
which originates from a neutralino mass insertion in the first
case. Due to the chirality flip, the fermion number conserving process
scales like $\Gamma \simeq m_{\tilde{f}}^3/m_{\tilde{\chi}^0_1}^2$
while the fermion number violating one scales like $\Gamma \simeq
m_{\tilde{f}}^5/m_{\tilde{\chi}^0_1}^4$ (If $f_i$ and $f_j$ had the
same chirality, the opposite situation would occur). 
Since $m_{\tilde{f}}
< m_{\tilde{\chi}^0_1}$, the fermion number conserving decay modes
tend to be dominant. 
A lower bound on the $\lambda$'s can be set in this case from the
requirement that  the two body R-parity violating 
decays are dominant, 
\begin{equation}
\lambda \simgt \alpha_w \frac{m_{\tilde{f}}}{m_{\tilde{\chi}^0}}.
\label{bound22}
\end{equation}
This lower bound tends
to be an order of magnitude lower than in the case of a real 
neutralino emission, discussed above. If the sfermion masses
are close to each other, 
the lower bound on the R-parity
violating coupling, Eq. (\ref{bound22}) may be even smaller 
due to a possible strong phase space suppression \cite{CGLW}.

In case {\it ii)} the dominant decay mode for non-NLSP sfermions is
R-parity conserving, overshadowing the R-parity violating decay modes.
Still, even for values of the R-parity violating couplings
$\lambda$'s below the  bounds, Eqs. (\ref{bound11}) and (\ref{bound22}),
R-parity violating interactions may  be detected via 
rare processes for a sizable range of values of those couplings.[9]

Until now, we have discussed the effects of trilinear couplings
in the superpotential.
What about R-parity violation by bilinears
in the scalar potential and, when are they present? 
In principle, new effects may appear in
this case, since the sneutrinos may acquire vacuum expectation 
values, inducing a mixing between the charged (neutral) leptons
and the charginos (neutralinos) of the  theory. The appearence of 
the bilinear terms in the scalar
potential may be due to the following
possibilities (we discuss here the scenario with explicit R-parity violation):
~\\
a) Only trilinear terms appear in the superpotential: In this
case, although absent at tree level, bilinear terms in the
scalar potential will be generated via the renormalization group
evolution of the scalar mass parameters. Indeed, 
in the presence of Higgs and 
R-parity violating Yukawa terms, the bilinear soft supersymmetry
breaking parameters are renormalized in a non-trivial 
way~\cite{WdeC}
\begin{equation}
\frac{d m_{H_1 L}^2}{d t} \simeq \frac{h_{1jk} \lambda_{Ljk}} {16 \pi^2}
\left(m_j^2 + m_k^2 + \frac{m_{H_1}^2 + m_L^2}{2} \right), 
\end{equation}
where $t = \ln (Q^2)$,  $m_{H_1 L}^2$ is the off-diagonal
Higgs slepton mass term appearing in the scalar potential, 
$h_{1 j k}$ 
are the Yukawa couplings of the
trilinear term $H_1 \Phi_i \Phi'_j$ in the superpotential, while
$\lambda_{Ljk}$ is the R-parity violating coupling of the 
trilinear term $L \Phi_i \Phi'_j$ in the superpotential. 
In the right hand side, we have only considered those terms 
proportional to the diagonal masses, which are presumably
dominant and we have ignored the contributions governed by the
soft supersymmetry breaking
trilinear mass terms.
The above means that in any theory in which trilinear couplings are
present at the GUT scale, even in the absence of bilinear 
terms in the superpotential, bilinear soft supersymmetry breaking
couplings will be generated at low energies.

Since the bilinear mass terms are only induced radiatively, with
effects proportional to the small Yukawa couplings of the theory,
we expect that the phenomenology will be still
dominated by the trilinear couplings, and not by the effects
of the small mixing. Neutrino masses are a possible exception
to this rule, and they provide the most efficient constraints
on the possible value of the sneutrino vacuum expectation values
(see discussion below). One should stress, however, that if the trilinear
R-parity violating couplings are allowed  and the supersymmetric
Higgs mass parameter $\mu \neq 0$, 
then there is no symmetry which
can protect against tree level R-parity violating 
bilinear terms in the scalar
potential and  in the superpotential. Therefore,
it is more realistic to consider case (b).

~\\
b) Both bilinear and trilinear  terms appear
in the superpotential:
It is convenient to work
in    the basis in which the $\epsilon_i L_i H_2$ terms, 
Eq.(\ref{eq:superpot}),   
have been rotated away
(for a basis independent approach see Refs. \cite{DE}) 
and to analyze the effects of the rotation. 
The new fields are then
given by,
\begin{equation}
H^{'}_1 = \frac{ \mu H_1 + \epsilon L}{\sqrt{\mu^2 + \epsilon^2}}
\;\;\;\;\;\;\;\;\;\;\;\;\;\;\;\;\;\;\;
L^{'} = \frac{ -\epsilon H_1 + \mu L}{\sqrt{\mu^2 + \epsilon^2}},
\end{equation}
where we have simplified the problem to the case of a single lepton
superfield.

In the  presence of soft supersymmetry breaking terms,
after rotation of the Higgs and lepton
superfields, these terms will generate not only 
additional trilinear couplings (see below),
but also bilinear terms in the scalar potential \cite{Hempfling},
\cite{NP},\cite{Nardi},\cite{DRV},
\begin{equation}
V(H_i,\tilde{L}) 
\simeq (m_{H_1}^2 - m_L^2) \sin( 2 \theta_{L})  H_1^{'} \tilde{L}^{'}
+ (B_H - B_L) \mu \sin\theta_{L} H_2 \tilde{L}^{'},
\label{scalpot} 
\end{equation}
where $\sin\theta_{L} = \epsilon/\sqrt{\mu^2 + \epsilon^2}$,
$m_{H_1}^2$ and $m_L^2$
are the scalar soft supersymmetry breaking parameters
for the Higgs and slepton fields, respectively, while
$B_H$ and $B_L$ are the soft supersymmetry breaking parameters
governing the proportionality of bilinear terms in the
potential and superpotential in the original basis.
%
%
Eq.(\ref{scalpot}) 
shows the dependence
of the bilinear R-parity violating terms in the scalar potential 
(generated by a bilinear term
in the superpotential)  at some
scale $M$ on the  soft terms at the same scale
$M$. Since bilinear terms generically generate vacuum expectation values
for sneutrinos, which in turn are strongly constrained by the limits
on the neutrino masses, eq.(\ref{scalpot}) is a strong constraint
on  the combination of R-parity violating $\epsilon_i$ couplings and on
the departure from universality of the soft supersymmetry breaking masses.
A detailed numerical analysis of this constraint is beyond the scope of this
paper. We observe only that
in the presence of 
universal soft supersymmetry breaking couplings at some scale,
the scalar potential has no bilinear R-parity 
breaking couplings at that scale. The kinetic terms for the Higgs
and neutrino fields are  a quadratic form and hence all what 
happens after the rotation is that we get a theory with new 
trilinear couplings $\bar{\lambda}_{ijk}$, with 
$\bar{\lambda} = \lambda$
for all couplings in which $L$ is not directly involved, while
\begin{equation}
\bar{\lambda}_{Ljk} = \lambda_{Ljk} + h_{1jk} \sin\theta_{LH}
\label{lambrot}
\end{equation}
where $h_{1jk}$ are the Yukawa couplings of the leptons and 
down quarks to the $H_1$ field. 
In this sense, in the presence of universal parameters at
some large scale, there is no difference between a theory
with only trilinear couplings, or a theory with both bilinear
and trilinear couplings.  
Indeed, 
due to the non-renormalization theorem, once rotated away,
there will be no regeneration of the
$\epsilon$ term at low energies.
In both cases, bilinear terms in 
the scalar potential will be generated radiatively at
low energies. 

The radiative generation of bilinear terms is  the condition
that one would like to achieve, in order to induce small neutrino
masses, which, as happens with the neutralino mass matrix in the
basis in which the $\epsilon$ parameters have been rotated away,
are proportional to the sneutrino vacuum expectation values
(the neutrino masses then still constrain the magnitude of the effective
trilinear couplings and of the soft masses). This
may happen in supergravity schemes, for exact universal boundary
conditions. This exact universality, although possible, is not
guaranteed by any symmetry in the theory. Another possibility
is that the $\mu$ and $\epsilon$ terms are generated at scales
larger than the scale leading to the superymmetry breaking
parameters. In that case, assuming that the dynamics leading
to supersymmetry breaking does not reintroduce them,
the $\epsilon$ terms may be rotated
away without worrying about the dangerous scalar terms in the
superpotential. 

A  possibility, more appropriate within the context of this
article is that supersymmetry is broken at scales much below
the GUT scale, through a gauge mediated mechanism, which assures
the universality of the Higgs and slepton soft supersymmetry
breaking mass parameters at the scale $M$, at which the 
messengers are decoupled. Assuming that the dynamics leading
to the appearence of the $\mu$, $\epsilon$ and $B$ terms
does not modify the soft supersymmetry breaking scalar mass
parameters, one can now rotate the $\epsilon$ parameter
away, generating only small bilinear soft terms through the
running between $M$ and $M_Z$. Observe that larger values of
the R-parity violating couplings will be allowed in this case,
due to the (assumed) small hierarchy of masses between $M$ and
$M_Z$.

Finally, one can address the question of the relative magnitude
of the two terms in Eq. (\ref{lambrot}). This is a model 
dependent question and in particular it is concievable that
additional (e.g. U(1)) symmetries broken at high scale, 
distinguish between $L$ and $H_1$ \cite{NP},  \cite{allignement}. 
Both the bilinear and
trilinear R-parity violating couplings are then realized as
non-renormalizable operators and the latter can be strongly
suppressed. Therefore, it is interesting to assume 
that there are regions where the effects
of the non-vanishing sneutrino vacuum expectation values 
are the dominant ones and
to see what kind of new phenomena they could
imply. In particular, we would like
to see the effects that can be obtained given the bounds on the
$\nu$ masses.

Under the above assumption, 
the bilinear R-parity violating
term in the scalar potential leads  
to a sneutrino vacuum expectation value and hence to
new patterns in the mass matrices for neutralinos
and charginos \cite{Valleetal}. 
Those states become now mixed with neutrinos and
charged leptons, respectively. In consequence, the structure of 
neutral and charged currents is modified and the couplings
$\tilde{\chi}^0_1 Z \nu$ and $\tilde{\chi}^0_1 l W$ ($\tilde{\nu} l l$ and $\tilde{l} \nu l$)
are generated at the tree level. 

At the tree-level we now have 
\begin{equation}
\tilde{\chi}^0_1 \rightarrow \nu Z \; \rightarrow \nu f \bar{f},
\end{equation}
with the width
\begin{equation}
\Gamma(\tilde{\chi}^0_1 \rightarrow \nu f \bar {f}) \simeq 2
((g_V^f)^2 + (g_A^f)^2)
\frac{G_F^2 M_{\tilde{\chi}^0_1}^5 \left(O^{``}_{\tilde{\chi}^0_1 \nu}
\right)^2}
{48 \pi^3} f(x),
\end{equation}
where $g_V^f$, $g_A^f$ are the vector and axial couplings of the
fermion $f$ to the $Z$ gauge boson and
\begin{equation}
O^{``}_{\tilde{\chi}^0_1 \nu} = \frac{1}{\sqrt{2}} 
\left[ N_{1 3} N_{l3} - N_{1 4} N_{l 4} 
+ \sum_{i=5}^7 N_{1 i} N_{l i} \right] \simeq g\frac{<\tilde{\nu}_L>}{M_2},
\end{equation}
where the 5th to 7th components of the neutralino mass matrix
in the basis of weak eigenstates are given by the neutrino fields,
and a non-negligible gaugino component of the neutralino has been assumed.
This coupling is suppressed by the mixing between the neutralino
and the neutrino, which is controlled by the sneutrino vacuum
expectation value, $<\tilde{\nu}_L>$ . The function $f(x)$ is a phase factor
given by \cite{Valleetal}
\begin{equation}
f(x) = \frac{12}{x^4} \left[ - \frac{x^2}{6} - \frac{1}{2}
+ \frac{1}{x^2} + \frac{1-x^2}{x^4} \ln\left(1-x^2\right) \right],
\end{equation}
with $x = M_{\tilde{\chi}^0_1}/M_Z$. 

The neutralino can also decay into $W^{+}$ and $l$, with a similar rate
to the one given above [20], 
\begin{equation}
\frac{\Gamma(\tilde{\chi}^0_1 \rightarrow f_u \bar{f}_d l)}
{\Gamma(\tilde{\chi}^0_1 \rightarrow
\nu f \bar{f})} \simeq 
\frac{ K_{\tilde{\chi}^0_1 l}^2}{8 \left((g_V^f)^2 + (g_A^f)^2 \right)
\left(O''_{\tilde{\chi}^0_1 l}\right)^2},
\end{equation}
with
\begin{equation}
K_{\tilde{\chi}^0_1 l}^2 \simeq  ( U_{l 2} N_{1 3} +\sqrt{2} U_{l 1}
N_{1 2} + \sum_{m=1}^{3} U_{l m+2} N_{1 m+4} )^2  +
( V_{l2} N_{1 4} - \sqrt{2}V_{l 1} N_{1 2} )^2,
\end{equation} 
where the 3rd to 5th components of the chargino mass matrices
are given by the neutrino fields, and we have ignored the difference
between the $Z$ and $W$ masses. 
It is easy to show that,
in the presence of a non-neglibile gaugino component
of the neutralino, $K_{\tilde{\chi}^0_1 l} \simeq g<\tilde{\nu}_L>/M_2$.

Comparing Eq. (2) with Eqs. (26) and (29), we conclude that 
the neutralino decay  into 
the $\gamma \tilde{G}$ will be the dominant channel if  
\begin{equation}
\frac{<\tilde{\nu}_L>}{M_2}\simlt \frac{10}{G_F F_{SUSY}}.
\label{bgmmix}
\end{equation}
For a characteristic value $\sqrt{F_{SUSY}}\approx 10^6$ GeV,
this bound is of order $10^{-6}$.
Hence, unless the mixing angle between the neutrinos and neutralinos
is extremely small, no photons will be observable in the presence
of the R-parity violating decays
induced by a sneutrino vacuum expectation value~\footnote{Observe that
photons may be induced by the additional R-parity violating channel,
$\tilde{\chi}^0_1 \rightarrow \gamma \nu$,
which is precisely the same signature as in gauge mediated models,
with the gravitino replaced by the neutrino. This 
radiative generated decay rate is, however,
subdominant compared to the three body decay channel}. 
Note, that the bound, Eq.(31), would be even more stringent
if a two body decay of the neutralinos into $Z \nu$ or $W l$ were open.

It is interesting to compare the above bounds, Eq. (\ref{bgmmix}),
with the experimental bounds coming from neutrino mass constraints.
A quantitative estimate of the bounds on the couplings and 
mixing angles may be obtained by considering the neutrino 
mass generation in cases a) and b), and assuming that
universal boundary conditions hold at some scale $\Lambda$.
The minimization of the scalar potential leads to
\begin{equation}
\frac{<\tilde{\nu}_L>}{\langle H_1 \rangle}
\approx \frac{m_{H_1 L}^2}{m^2_{\tilde{\nu}_L}},
\end{equation}
where $m_{\tilde{\nu}_L}$ is the running sneutrino mass.
Under the assumption of radiative generation of
the bilinear terms, Eq. (32) can be rewritten as
\begin{equation}
<\tilde{\nu}_L> \simeq 
\frac{\lambda_{L jk} \langle H_1 \rangle h_{1 jk}}{16 \pi^2}
\ln \left( \frac{\Lambda^2}{M_Z^2} \right),
\label{eq:33}
\end{equation}
where we are ignoring factors of order one, proportional
to the mass hierarchy of different sparticles (Higgs) at
low energies \cite{Nardi}. 
{}From the neutralino mass matrix, it appears that the mass
of the neutrinos is of order
\begin{equation}
m_{\nu} \simeq g^2\frac{<\tilde{\nu}_L>^2}{M_2}, 
\end{equation}
where $M_2$ is a typical gaugino mass, of order of the
weak scale. Hence, as a function of the experimental bound on the neutrino
masses, $m_{\nu}^{\rm exp}$, 
strong constraints on the sneutrino vaccum expectation
values, (and hence on the couplings and mixings) appear, 
\begin{equation}
g<\tilde{\nu}_L> \simlt \sqrt{M_2 m_{\nu}^{\rm exp}}.
\end{equation}
For the three neutrinos we have~ \cite{PDG}: 
$m_{\nu_{\tau}}^{\rm exp} \simlt 24 MeV$, 
$m_{\nu_{\mu}}^{\rm exp} 
\simlt 170 KeV$, and $m_{\nu_{e}}^{\rm exp} \simlt 15 eV$.
For the  second and third generations the bound becomes even stronger if
the cosmological constraints coming from avoiding
overclosing the universe~\cite{Raffelt} and valid for any generation,  
are considered,
\begin{equation}
m_{\nu} \simlt {\rm 40 eV}.
\end{equation}
Therefore, 
the sneutrino vacuum expectation value is constrained to be below
\begin{equation}
\frac{<\tilde{\nu}_L>}{M_2} \simlt 10^{-5},
\;\;\;\;\;\; {\rm or, equivalently~~~ for~~~} M_2 \simeq M_Z 
\;\;\;\;\;\;\;
<\tilde{\nu}_L> \simlt 1 MeV.
\label{eq:37}
\end{equation}
We see that the cosmological bounds on the sneutrino vacuum expectation value
are close to the  upper bound, Eq. (\ref{bgmmix}),
leaving only a small window for the R-parity violating decays to overshadow
the decays into gravitinos.\footnote{A bound 
on the R-parity violating Yukawa couplings may also be obtained from 
Eqs. (\ref{eq:33}) and (\ref{eq:37}).
Going to the basis in which the down quark
and lepton masses are diagonal, the sneutrino vacuum
expectation value becomes proportional to the square of those
masses. Hence, a bound on the R-Parity violating coupling
may be obtained from these considerations:
\begin{equation}
\lambda_{Ldd}^2 \simlt \frac{m_{\nu}^{exp}M_2}{m_d^2 (N_C^d)^2} 
\left(\frac{16 \pi^2}{\log(\Lambda^2/M_Z^2)}\right)^2
\end{equation}
where $m_d$ is the mass of the lepton or
quark and $N_C^d$ is it number of colors. This means that
the coupling $\lambda_{Lbb}$ is heavily constrained by
these considerations. Independently of the value of $\Lambda$,
it follows $\lambda_{Lbb} \simlt 10^{-3}$.
This bound is of the same order, but
somewhat more general than the usual bound obtained
from the low energy radiative corrections to neutrino 
masses \cite{rpreview},
since it does not depend on the exact value of the left-right
mixing mass parameter.}
Observe that, avoiding 
the above bound would require the existense of
new, non-radiative decay modes of the heavy neutrinos, what
would imply going beyond the physical framework under 
discussion~\cite{Raffelt}. 
For instance, 
the cosmological bounds may be avoided in the presence of
Majorons, as those appearing in models of spontaneous breaking
of R-parity \cite{Valleetal}, 
which we shall not discuss in this paper. 

Other processes which can take place, via the neutralino-neutrino 
mixing are
\begin{equation}
\tilde{\nu}_L \rightarrow l_i \bar{l}_i  \;\;\;\;\;\;\;\;\;\;\;\;
\tilde{l} \rightarrow l_i \bar{\nu}_i 
\end{equation}
with  widths
\begin{equation}
\Gamma \simeq \left(\frac{g^2 <\tilde{\nu}_L>}{M_2} \right)^2 
\frac{m_{\tilde{\nu}}}{16\pi} \;\;\;\;\;\;\;\; {\rm and}
\;\;\;\;\;\;\;\;\;
\Gamma \simeq \left(\frac{g^2 <\tilde{\nu}_L>}{M_2} \right)^2 
\frac{m_{\tilde{l}}}{16\pi},
\end{equation}
respectively, where for simplicity we have ommitted factors of order one.
The decay mode of sleptons and sneutrinos is induced by 
gauge coupling interactions.  
These processes will look similar to those
induced via trilinear R-parity violating couplings. 
Hence, for the gravitino decay mode to be the dominant one, the bound
on the sneutrino vacuum expectation value is defined 
by Eq. (\ref{bound2}), with $\lambda$  replaced by 
$g^2 <\tilde{\nu}_L>/M_2$.

A final comment is in order. As said in the introduction, 
if the supersymmetry breaking scale is large, photons
are not a characteristic signature of the minimal supersymmetric
standard model, unless specific conditions are fulfilled. This can
happen, for instance, when $M_2 \simeq M_1$ and $\tan\beta \simeq 1$, in
which case the two lightest eigenstates 
are a photino and a Higgsino~\cite{m2eqm1}.
For low values of $\mu$ one can get the Higgsino as the LSP, and
the photino will have a one-loop induced decay into a Higgsino 
and a photon
\begin{equation}
\tilde{\gamma} \rightarrow \gamma + \tilde{H}
\end{equation}
with a rate 
\begin{equation}
\Gamma \simeq \frac{G_F \alpha^2}{(4 \pi)^3} 
\left[ \frac{m_{\tilde{\gamma}}^2 - m_{\tilde{H}}^2}
{m_{\tilde{\gamma}}} \right]^3 G(m_{\tilde{\chi}^+}),
\label{phohiggamma}
\end{equation}
where $G(m_{\tilde{\chi}^+}) \simeq {\cal O}(m^2_W/m^2_{\tilde{\chi}^+})$.
In general, the actual rate will be given by the above expression,
but replacing $G(m_{\tilde{\chi}^+})$ by a complicated function of all
particles contributing to the loop. In the above, we have assumed
that the chargino-$W^+$ contribution is the dominant one.
The photon signatures arising from these models are
easily distinguishable from the ones arising in 
low energy supersymmetry breaking models by the presence of massive final
states carrying missing energy. In addition, the rate,
Eq. (\ref{phohiggamma}), is not suppressed by any strong factor
and hence, in case it becomes the relevant one, it will overshadow
the R-parity violating decays of the neutralinos for values of
the trilinear couplings $\lambda \simlt 10^{-2}$.

We conclude that the observation of signatures with photons and 
missing energy, will 
put very  strong limits on all the trilinear R-parity violating
couplings appearing in the superpotential (or bilinear couplings 
in the  cases that their effects are dominant compared
to those induced by the trilinear couplings). 
The only exception to these rules can occur in 
the heavy gravitino  case, for particular conditions in which
the decay of photinos into a photon and Higgsinos is not suppressed.
This decay process is easily distinguishable from the decay into LSP
gravitinos. Relatively large
R-parity violating couplings may still be present
in this case. In the case of bilinear R-parity violating terms
in the superpotential,
we have obtained strong bounds on the sneutrino
vacuum expectation values, and hence on the possible neutrino--neutralino
or lepton chargino mixing coming from the constraints on
the neutrino masses. If the cosmological bounds are considered,
there is only a small window of allowed neutrino-neutralino (or lepton
-chargino) mixings, 
which would
allow the observation of mixing-induced R-parity violating decays
in models of low energy supersymmetry breaking. 
~\\
~\\
~\\
{\bf ACKNOWLEDGEMENTS}
~\\
M.C. and C.W. would like to thank the members of the theory
group at the University of Buenos Aires, Argentina, for their
kind hospitality during the completion of this work.
S.P. is partially supported by US-Poland  Maria Sklodowska-Curie Joint Fund II (grant
MEN/DEO-96-264) and by the Polish State Committee for Scientific Research 
under the
grant 2 P03B 040 12. The authors would
like to thank the Aspen Center for Physics, where most of this
work has been done.

\end{document}